\documentclass{amsart}
\usepackage{amssymb}
\usepackage{graphicx}
\usepackage{amscd}
\usepackage{color}

\newtheorem{theorem}{Theorem}
\theoremstyle{plain}

\newtheorem{remark}{Remark}

\numberwithin{equation}{section}

\begin{document}
\title[Representation Theorems for Poverty indices]{Asymptotic
Representation Theorems for Poverty Indices}
\author[G. S. Lo]{Gane Samb LO}
\address{LSTA, Universit\'{e} Paris VI. FRANCE and LERSTAD, Universit\'{e}
Gaston Berger, SENEGAL.}
\email{ganesamblo@ufrsat.org}
\author[S. T. Sall]{Serigne Touba Sall}
\address{FASTEF, Universit\'{e} Cheikh Anta Diop and LERSTAD, Universit\'e
Gaston Berger de Saint-Louis.}
\email{stsall@ufrsat.org}

\begin{abstract}
\large
We set general conditions under which the general poverty index, which
summarizes all the available indices, is asymptotically represented with
some empirical processes. This representation theorem offers a general key,
in most directions, for the asymptotics of the bulk of poverty indices and
issues in poverty analysis. Our representation results uniformly hold on a
large collection of poverty indices. They enable the continuous measure of
poverty with longitudinal data.
\end{abstract}

\maketitle

\Large

In quantitative poverty analysis, poverty indices are the key tools as well
as inequality measures. A great number of such indices have been introduced
in the literature since the pioneering works of the Nobel Prize winner,
Amartya Sen (1976) who first derived poverty measures (see %
\textcolor{red}{\cite{sen}}) from an axiomatic point of view. A survey of
these indices is to be found in Zheng \textcolor{red}{\cite{zheng}}, who
also discussed their properties and classified them from an axiomatic point
of view.\newline

Statistical asymptotic laws for these indices, particularly asymptotic
normality, on which statistical inference on the unknown poverty index may
be based, are also of great importance. Recent works which dealt with this,
are available in \textcolor{red}{\cite{barrett}}, \textcolor{red}{%
\cite{duclosdav}}, \textcolor{red}{\cite{bfzheng}} and \textcolor{red}{%
\cite{bchowzheng}} for instance. These results reveal themselves very
powerful and showed real interest in applications. Nevertheless, the indices
are studied mostly one by one. In \textcolor{red}{\cite{lo2}}, a unified
approach is proposed with a general form of the poverty indices, named the
General Poverty Index (GPI), including almost all the proposed indices. In %
\textcolor{red}{\cite{lo2}} and \textcolor{red}{\cite{lo3}}, a general
asymptotic theory of the GPI is given, based on the so-called Hungarian
approximations (see \textcolor{red}{\cite{cchm}} and \textcolor{red}{%
\cite{ch}}). Now there is much to do when we deal with longitudinal data. In
this case, one has to move from a static approach to a time-dependent one.
Moreover, the GPI and a large class of inequality measures, form classes of
L-Statistics indexed by functions.\\

\noindent We aim at giving general tools from which, the functional and time-dependent
asymptotic laws of the GPI will be derived. Precisely, we give here
functional and time-dependent representation theorems of the class of
poverty indices into a functional empirical process, largely described in %
\textcolor{red}{\cite{vaart}}, and a new functional process to be especially
handled in \textcolor{red}{\cite{lo1}}.\\

\noindent Now, let us make some notation that will permit to clarify the ideas and
formulate the problems. We consider a population of individuals, each of
which having a random income or expenditure $Y$, with distribution function $%
G$. An individual or a household is considered as poor whenever $Y$ fulfills 
$Y\leq Z$, where $Z>0$ is a specified theshold level named the poverty line.\\

\noindent Consider now a random sample $Y_{1},Y_{2},...Y_{n}$\ of size $n$\ of
incomes, with empirical distribution function $G_{n}(y)=n^{-1}\#\left\{
Y_{i}\leq y:1\leq i\leq 1\right\} $. The number of poor individuals within
the sample is then equal to $Q_{n}=nG_{n}(Z)$. Let also be measurable
functions $A(p,q,z)$, $w(t)$, and $d(t)$ of $p,q\in \mathbb{N},$\ and $%
z,t\in \mathbb{R}$ and $B(q)=\sum_{i=1}^{q}w(i).$\\

\noindent Let finally $Y_{1,n}\leq Y_{2,n}\leq ...\leq Y_{n,n}$\ be the order
statistics of the sample $Y_{1},Y_{2},...Y_{n}$\ of $Y$. We introduce the
following

\begin{equation}
J_{n}=\frac{A(Q_{n},n,Z)}{nB(Q_{n})}\sum_{j=1}^{Q_{n}}w(\mu _{1}n+\mu
_{2}Q_{n}-\mu_{3}j+\mu_{4})\: d\left( \frac{Z-Y_{j,n}}{Z}\right),
\label{ssl01}
\end{equation}

\noindent where $\mu _{1},\mu _{2},\mu _{3},\mu _{4}$\ are constants, as the
General Poverty Index (GPI) like in \textcolor{red}{\cite{lo2}} and %
\textcolor{red}{\cite{lo3}}.\\

\noindent We already showed in \textcolor{red}{\cite{lo3}} how to derive \ from %
\textcolor{blue}{(\ref{ssl01})} the individual poverty measures like the
Sen, Kakwani, Foster-Greer-Thorbecke, Thon, Chakravarty ones and some other
ones. (See \textcolor{red}{\cite{zheng}}, for a definition of such
measures). We do not need to return back to this in this paper.

As said previously, our aim is to obtain functional asymptotic laws for the
time-dependent GPI. We then need to define our index set in %
\textcolor{blue}{(\ref{ssl01})}. Suppose that the functions $A$, $w$ and $d$
are in some classes of positive and measurable functions $\mathcal{C}%
_{i},i=1,2,3$\ with these further specifications : $\mathcal{C}_{1}$\ is a
class of functions $A(p,q,z)$ with $(p,q,z)\in \mathbb{N}^{3},$ $\mathcal{C}%
_{2}$ of functions $w(t)$\ with $t\in \mathbb{R}$\ and $\mathcal{C}_{3}$ of
functions $d(\cdot )$ continuous and defined on $[0,1]$ onto $[0,1]$ and
bounded by one. The constants vector $\mu =(\mu _{1},...,\mu _{4})^{t}$\
lies also in some subset $\mathcal{C}_{4}$ of $\mathbb{N}^{4}$. We put $%
\lambda_{0} =(A,w,u)\in \Gamma{0} =\mathcal{C}_{1}\mathcal{\times C}_{2}\mathcal{%
\times C}_{4}.$\ In the longitudinal data case, we observe the same
households over the time. This leads to the longitidunal observations of $%
Y\in C([0,T]),$
\begin{equation*}
\{Y_{1}(t),...,Y_{t}(t),0\leq t\leq T\},
\end{equation*}%
where for each $t\in \lbrack 0,T]$, $G_{t}(\cdot )$\ stands for the
distribution function of $Y(t)$. We consequently use the index 
\begin{equation*}
\lambda =(A,w,u,t)\in \Gamma =\mathcal{C}_{1}\mathcal{\times C}_{2}\mathcal{%
\times C}_{4}\times \lbrack 0,T]
\end{equation*}%
and%
\begin{equation*}
\phi =(A,w,d,u,t)\in \Phi =\mathcal{C}_{1}\mathcal{\times C}_{2}\mathcal{%
\times C}_{3}\times \mathcal{C}_{4}\times \lbrack 0,T].
\end{equation*}%
In the time-dependent case, the poverty line $Z(t)$ may depend on the time
and so does the poor headcount denoted as $Q_{n}(t).$ With these notations,
our object study becomes%
\begin{equation*}
J_{n}(\phi )=\frac{A(Q_{n}(t),n,Z(t))}{nB(Q_{n}(t))}\sum_{j=1}^{Q_{n}(t)}w(%
\mu _{1}n+\mu _{2}Q_{n}(t)-\mu _{3}j+\mu _{4})\text{ } d\left( \frac{%
Z(t)-Y_{j,n}(t)}{Z(t)}\right).
\end{equation*}%

\bigskip

\noindent We are now giving our foundamental result, that is the uniform
representation of $\{\sqrt{n}(J_{n}(\phi )-J(\phi )),\phi \in \Phi \}$ in
terms of functional empirical processes, where $J(\phi )$ is the exact GPI.\\

\noindent In the sequel, we use almost sure limits in outer probability $(a.s.o.p)$,
limits to zero in outer probability denoted $o_{P}^{\ast }(1)$, following
the new theory in \textcolor{red}{\cite{vaart}}, when dealing with
non-measurable processes. We point out for once that all the limits in this paper are meant as $n\rightarrow +\infty$, unless the contrary is specified.\\

\section{Fundamental Theorem}

\label{sec2}

\noindent Define $\mathcal{F}_{0}=\{f_{t}:x\leadsto 1_{(x(t)\leq Z(t)},t\in
\lbrack 0,T]\}$ as a subset of $\ell ^{\infty }(C([0,T])$, the set of real
bounded and continuous functions defined on $C([0,T])$. We will consider
these general assumptions.

\begin{itemize}
\item[(HP0)] There exist $\beta >0$ and $0<\xi <1$ such that 
\begin{equation*}
0<\beta <\inf_{0\leq t\leq T}G_{t}(Z)<\sup_{0\leq t\leq T}G_{t}(Z)<\xi <1.
\end{equation*}

\item[(HP1)] The familiy $\mathcal{F}_{0}$ is a $\mathbb{P}_{Y}-$%
Glivenco-Cantelli class, that is, as $n\rightarrow \infty ,$%
\begin{equation*}
\sup_{t\in \lbrack 0,T]}\left\vert G_{t,n}(Z(t))-G_{t}(Z(t))\right\vert
\rightarrow 0,\text{ }a.s.o.p.
\end{equation*}%
where, for any $t\in \lbrack 0,T]$ and $y\in \mathbb{R}$, $%
G_{t,n}(y)=n^{-1}\sum_{i=1}^{n}1_{(Y(t)\leq y)}$.

\item[(HP2a)] There exist a class $\mathcal{C}$ of functions $c(s,t)$ of $%
(s,t)\in \mathbb{R}^{2}$ and a class $\Pi $ of functions $\pi (s,t)$ of $%
(s,t)\in \mathbb{R}^{2},$ such that for any $\lambda =(A,w,\mu ,t)\in \Gamma
,$ one can find a single function $h(p,q)$ of $(p,q)\in \mathbb{N}^{2}$ such
that there exist a function $c\in \mathcal{C}$ and a function $\pi \in \Pi $%
, all of them independant of $t\in \lbrack 0,T\rbrack$, under which we have,
as $n\rightarrow +\infty $, 
\begin{equation*}
\sup_{\lambda \in \Gamma }\max_{1\leq j\leq Q_{n}(t)}\left|
A(n,Q_{n}(t))h^{-1}(n,Q_{n}(t))w(\mu _{1}n+\mu _{2}Q_{n}(t)-\mu _{3}j+\mu
_{4})\right.
\end{equation*}
\begin{equation*}
\left. -c(Q_{n}(t)/n,j/n)\right| =o_{P}^{\ast }(n^{-1/2}).
\end{equation*}

\item[(HP2b)] and 
\begin{equation*}
\sup_{\lambda \in \Gamma }\max_{1\leq j\leq Q_{n}(t)}\left|
w(j)h^{-1}(n,Q_{n}(t))-\frac{1}{n}\pi (Q_{n}(t)/n,j/n)\right|
=o_{P}^{\ast}(n^{-1/2})
\end{equation*}
\item[(H2Pc)]  There exists a function  $c(u,v)$  of $(u,v)\in (0,1)^{2}$
independent of $t\in \lbrack 0,T],$ such that, as $n\rightarrow +\infty $, 

\begin{equation*}
\sup_{t\in \lbrack 0,T]}\max_{1\leq j\leq Q_{n}(t)}\left|
A(n,Q_{n}(t))h^{-1}(n,Q_{n}(t))w(\mu _{1}n+\mu _{2}Q_{n}(t)-\mu _{3}j+\mu
_{4})\right. 
\end{equation*}
\begin{equation*}
\left. -c(Q_{n}(t)/n,j/n)\right| =o_{P}^{\ast }(n^{-1/2}).
\end{equation*}

\item[(HP3)] The elements of the classes $\mathcal{C}$ and $\Pi $ have
equi-continuous partial differentials in the sense that (for example for $c
\in \mathcal{\mathcal{C}}$) : 
\begin{equation*}
\lim_{(k,l)\rightarrow (0,0)}\sup_{(x,y)\in (0,1)^{2}}\sup_{c\in C}\left| 
\frac{\partial c}{\partial y}(x+l,y+k)-\frac{\partial c}{\partial y}%
(x,y)\right| =0,
\end{equation*}

and 
\begin{equation*}
\lim_{(k,l)\rightarrow (0,0)}\sup_{\beta \leq x \leq \xi, \text{} y\in
(0,1), }\sup_{c\in C}\left| \frac{\partial c}{\partial x}(x+l,y+k)-\frac{%
\partial c}{\partial x}(x,y)\right| =0.
\end{equation*}

\item[(HP4)] For any $(c,\pi )\in \mathcal{C}\times \Pi ,$ for fixed $x$,
the functions $y\rightarrow \frac{\partial c}{\partial y}(x,y)$ and $%
y\rightarrow \frac{\partial \pi }{\partial y}(x,y)$ are monotone.

\item[(HP5)] For any $t \in \lbrack0,T \rbrack$, $G_{t}$ is strictly
increasing, and the functions $G_{t}$ are equi-continuous in $t\in \lbrack
0,1\rbrack$.

\item[(HP6)] There exist $H_{0}>0$ and $H_{\infty }<+\infty $ such that, for
any $(t,c,\pi ,d)\in \lbrack 0,T]\times \mathcal{C}\times \Pi \times 
\mathcal{C}_{3},$\newline
\begin{equation*}
H_{0}<H_{c}(\phi )=\int_{0}^{+\infty} c(G_{t}(Z),G_{t}(y))\gamma
(y)dG_{t}(y)<H_{\infty },
\end{equation*}
\noindent and 
\begin{equation*}
H_{0}<H_{\pi }(\phi )=\int_{0}^{+\infty} \pi
(G_{t}(Z),G_{t}(y))e(y)dG_{t}(y)<H_{\infty };
\end{equation*}

\noindent where $\gamma (x)=d(\frac{Z-x}{Z})1_{(x\leq Z)}$ and $%
e(x)=1_{(x\leq Z)}$ for $x\in \mathbb{R}$. Here and in the sequel, the
functions depend, in some way, on $\phi$ even when we do not specify it or
even when we only partially do it.

\item[(HP7)] There is a universal constant $K_{0,}$ such that there exists $%
\delta >0$, there exists $r>0$ such that 
\begin{equation}
\left| s-t\right| \leq \delta \Longrightarrow \left| \frac{1}{3}-\mathbb{E}%
(G_{t}(Y(t))G_{s}(Y(s))\right| \leq K_{0}\left| s-t\right| ^{1+r},
\label{unif01}
\end{equation}
\end{itemize}

\noindent Here is our fondamental tool Theorem.

\begin{theorem}
\label{theo1} Suppose that the hypotheses (HP1)-(HP7) hold. Put 
$$J(\phi)=H_{c}(\phi )/H_{\pi }(\phi ),
$$
\begin{equation*}
g_{t}=H_{2}^{-1}g_{c}-H_{c}H_{\pi}^{-2}g_{\pi }+Ke(f_{t}(\cdot ))
\end{equation*}%
with 
\begin{equation}
g_{c}(\cdot )=c(G_{t}(Z),G_{t}(f_{t}(\cdot )))\gamma (f_{t}(\cdot )),\text{ }%
g_{\pi }=\pi (G_{t}(Z),G_{t}(f_{t}(\cdot )))e(f_{t}(\cdot )),  \label{meth2}
\end{equation}%
\begin{equation}
K(\phi )=H_{\pi}^{-1}K_{c}-H_{c}H_{\pi}^{-2}K_{\pi }  \label{meth3}
\end{equation}

\begin{equation}
K_{c}(\phi )=\int_{0}^{1}\frac{\partial c}{\partial x}(G_{t}(Z),s)\gamma
(G_{t}^{-1}(s))ds,\text{ }K_{\pi }(\phi )=\int_{0}^{1}\frac{\partial \pi }{%
\partial x}(G_{t}(Z),s)e(G_{t}^{-1}(s))ds,  \label{meth4}
\end{equation}
\noindent 
\begin{equation*}
\nu =H_{\pi}^{-1}\nu _{c}-H_{c}H_{\pi}^{-2}\nu _{\pi },
\end{equation*}
\noindent where 
\begin{equation*}
\nu _{c,t}(y)=\frac{\partial c}{\partial x}(G_{t}(Z),G_{t}(f_{t}(y)))%
\gamma(f_{t}(y)), \: \nu _{\pi,t}(y)=\frac{\partial \pi}{\partial x}%
(G_{t}(Z),G_{t}(f_{t}(y))) e(f_{t}(y)).
\end{equation*}

\noindent Then we have, uniformly in 
\begin{equation*}
\phi = (A,w,d,\mu ,t)\in \Phi =\mathcal{C}_{1}\times \mathcal{C}%
_{2}\times \mathcal{C}_{3}\times \mathcal{C}_{4}\times \lbrack 0,T],
\end{equation*}%
the following representation : 
\begin{equation}
\sqrt{n}(J_{n}(\phi )-J(\phi ))=\alpha _{t,n}(g_{t})+\beta _{n}(t,\nu
_{t})+o_{P}^{\ast }(1),  \tag{R}
\end{equation}%

\noindent with

$$
\alpha _{t,n}(g_{t})=\frac{1}{\sqrt{n}} \sum_{j=1}^{n} g_{t}(Y_{j}) - \mathbb{E}g(Y_{j})
$$

\noindent and

\begin{equation*}
\beta _{n}(t,\nu _{t})=\frac{1}{\sqrt{n}}\sum \left\{
G_{t,n}(Y_{j}(t))-G_{t}(Y_{j}(t))\right\} \nu _{t}(Y_{j}).
\end{equation*}
\noindent Suppose that (H2Pc) holds in place of (H2Pa) and (HP2b) and the other assumptions atre true. Then the representation (R) holds with

\begin{equation}
K(t)=K_{c}(t),g_{t}=g_{c,t}\text{ and }v_{t}=\nu _{c,t} \tag{RD}
\end{equation}
\end{theorem}

\begin{remark}
The conditions may seem hard to hold. But, for the classical poverty
measures, (HP2a), (HP2b) and (HP4) hold. (HP4$)$ also holds for the Kakwani
class. The other hypotheseses depend on the distribution of $\{Y(t),t\in
\lbrack 0,T]\}$ and the properties of the poverty lines $\{Z(t),t\in \lbrack
0,T]\}$.
\end{remark}

\begin{remark}
With the representation $(R)$, $\sqrt{n}(J_{n}(\phi )-J(\phi ))$ may be
studied in many ways whenener the properties of the processes $\alpha
_{t,n}(g_{t})$ and $\beta _{n}(t,\nu _{t})$ are known as well as their
covariance structure. The first is nothing else than the functional process.
It is largely studied in \textcolor{red}{\cite{vaart}}. The second $\beta
_{n}(t,\nu _{t}),$ apparently new, is entirely described in %
\textcolor{red}{\cite{lo1}} as well as its correlation with the functional
empirical process. Thus, Formula (R) opens various poverty study fields :
using poverty indices with longitudinal data, simultaneous comparison poverty
situations with several indices, statistical estimation of lack of
decomposability, uniform boostraping of poverty indices, etc. Such studies
using $(R)$ are underway. As an example, it is used in \textcolor{red}{%
\cite{lo4}} to get confidence intervals of the relative change in poverty.
This method enables to check whether the Millenium Development Goal (MDG) of
halving poverty in some time interval is achieved with a probability
confidence, say 95\%. In the same time, it is showed in \textcolor{red}{%
\cite{lo4}} how to combine $(R)$ with the results of $\beta _{n}(t,\nu _{t})$
in \textcolor{red}{\cite{lo1}} to get concluding applications in specific
problems.
\end{remark}

\section{PROOFS}

\label{sec3}

\noindent Let us begin by giving general considerations. We introduce these
notations to be used later in the proofs. First, based on the hypothesis $(H5)
$, we may use the following representation. First, define the rank statistic 
$R_{n}(t)=(R_{1,n}(t),...,R_{n,n}(t))$ based on $Y_{1}(t),...,Y_{n}(t)$
defined by 
\begin{equation*}
\forall (i,j)\in \{1,...,n\}^{2},\text{ }R_{j,n}(t)=i\Leftrightarrow
Y_{j}(t)=Y_{i,n}(t).
\end{equation*}

\noindent Since each $G_{t}$ is a one-to-one mapping function, we have 
\begin{equation*}
G_{t,n}(Y_{j})=R_{j,n}(t)/n,\text{ a.s.}
\end{equation*}

\noindent Secondly, we remark that $%
U_{1}(t)=G_{t}(Y_{1}),U_{2}(t)=G_{t}(Y_{2}(t)),...$ are independent uniform
random variables whenever G$_{t}$ is increasing. We will have to consider $%
U_{t,n}(\cdot)$ and $V_{t,n}(\cdot)$, respectively the empirical
distribution and quantile functions based on $U_{1}(t),...,U_{n}(t)$. Denote
also $\alpha_{t,n}(s)=\left\{ \sqrt{n}(U_{t,n}(s)-s), \text{ } s\in (0,1)
\right\}$, the empirical process based on $Y_{1}(t),Y_{2}(t),...,Y_{n}(t)$;
for $n\geq 1$.

\noindent Now, by (HP1), 
\begin{equation*}
\lim_{n\rightarrow \infty }\sup_{t\in \lbrack 0,T]}\sup_{y\in \mathbb{R}%
}\left\vert G_{t,n}(y)-G_{t}(y)\right\vert =0,\text{ }a.s.o.p
\end{equation*}%
By $(HP2)$, uniformly in $\phi \equiv (A,w,\mu ,d,t)\in $ $\Phi =C_{1}\times
C_{2}\times C_{3}\times C_{4}\times \lbrack 0,T],$%
\begin{equation*}
J_{n}=\frac{A(n,Q_{n}(t))/h(n,Q_{n}(t),j)}{%
n\sum_{j=1}^{Q_{n}(t)}w(j)/h(n,Q_{n}(t),j)}\sum_{j=1}^{Q_{n}(t)}w(\mu
_{1}n+\mu _{2}Q_{n}(t)-\mu _{3}j+\mu _{4})d(\frac{Z-Y_{j,n}(t)}{Z}).
\end{equation*}%
\begin{equation}
=(J_{n}(c)+o_{P}^{\ast }(n^{-1/2})/(J_{n}(\pi)+o_{P}^{\ast }(n^{-1/2}),
\label{p01}
\end{equation}%
where

\begin{equation}
J_{n}(c)=\frac{1}{n}\sum_{j=1}^{n}c(G_{t,n}(Z),G_{t,n}(Y_{j}))\gamma
(Y_{j}(t)),  \label{p02}
\end{equation}
and 
\begin{equation}
J_{n}(\pi )=\frac{1}{n}\sum_{j=1}^{n}\pi (G_{t,n}(Z),G_{t,n}(Y_{j}))e(Y_{j}).
\label{p03}
\end{equation}
Now we have

\begin{equation*}
J_{n}(c)=\int_{0}^{1}c(G_{t,n}(Z),U_{t,n}(V_{t;n}(s)))\gamma
(G^{-1}(V_{t,n}(s)))ds,
\end{equation*}
and 
\begin{equation*}
H_{c}(\phi )=\int_{0}^{1}c(G_{t}(Z),s)\gamma (G^{-1}(s))ds.
\end{equation*}
Then 
\begin{equation*}
J_{n}(1)=\frac{1}{n}\sum_{j=1}^{n}c(G_{t}(Z),G_{t}(Y_{j}))\gamma (Y_{j}(t))
\end{equation*}
\begin{equation*}
+\frac{1}{n}\sum_{j=1}^{n}\left\{
c(G_{t,n}(Z),G_{t,n}(Y_{j}))-c(G_{t}(Z),G_{t,n}(Y_{j}))\right\} \gamma
(Y_{j}(t))
\end{equation*}
\begin{equation*}
+\frac{1}{n}\sum_{j=1}^{n}\left\{
c(G_{t}(Z),G_{t,n}(Y_{j}))-c(G_{t}(Z),G_{t}(Y_{j}))\right\} \gamma
(Y_{j}(t)).
\end{equation*}
Set, 
\begin{equation*}
D_{1}=\frac{1}{n}\sum_{j=1}^{n}\left\{
c(G_{t,n}(Z),G_{t,n}(Y_{j}))-c(G_{t}(Z),G_{t,n}(Y_{j}))\right\} \gamma
(Y_{j}(t)),
\end{equation*}
and 
\begin{equation*}
D_{2}=\frac{1}{n}\sum_{j=1}^{n}\left\{
c(G_{t}(Z),G_{t,n}(Y_{j}))-c(G_{t}(Z),G_{t}(Y_{j}))\right\} \gamma
(Y_{j}(t))\equiv D_{1}+D_{2}.
\end{equation*}

\noindent First, we have by using the classical representations, 
\begin{equation*}
D_{1}=\int_{0}^{1}\left\{ c(G_{t,n}(Z),U_{t,n}(V_{t,n}(s)))-c(G_{t}(Z),U_{t,n}(V_{t;n}(s)))\right\}\gamma
(G^{-1}(V_{t,n}(s)) ds
\end{equation*}
\begin{equation}
=(G_{t,n}(Z)-G_{t}(Z))\int_{0}^{1}\left\{ \frac{\partial c}{\partial x}%
(\zeta _{n}(t,Z),U_{t,n}(V_{t;n}(s))\gamma (G^{-1}(V_{t,n}(s))\right\} ds
\label{methodn}
\end{equation}
where $\zeta _{n}(t,Z)$ lies between $G_{t}(Z)$ and $G_{t,n}(Z)$. We want
now to prove that integral factor in \textcolor{blue}{(\ref{methodn})} tends
to $K_{c}(\phi )$ uniformly in $\phi $. We shall use a method that will be
repeated throughout the paper. First, we remark that this is performed in $%
s\in (V_{t,n}(s)\leq G_{t}(Z))$. Since $V_{t,n}(s)\rightarrow s$ uniformly
in $(t,s)\in (0,T)\times (0,1)$, in outer probability, we have for any $%
\epsilon >0, 0<\xi +\epsilon <1$, $(V_{t,n}(s)\leq G(Z))\subseteq (s\leq
G_{t}(Z)+\epsilon )$, uniformly in $(t,s)\in (0,T)\times (0,1),$ with outer
probability greater or equal to $1-\epsilon $ (denoted $w.p.1-\epsilon $),
for large values of n. Thus for such n's, we have, uniformly in $(t,s)\in
(0,T)\times (0,1),$ $w.p.1-\epsilon $, 
\begin{equation*}
\left| \int_{0}^{1}\frac{\partial c}{\partial x}(\zeta
_{n}(t,Z),U_{t,n}(V_{t,n}(s))\gamma (G_{t}^{-1}(V_{t,n}(s)))ds\right.
\label{method1}
\end{equation*}
\begin{equation*}
\left. -\int_{0}^{1}\frac{\partial c}{\partial x}(G_{t}(Z),s)d(\frac{%
Z-G_{t}^{-1}(s)}{Z})e(G_{t}^{-1}(V_{n}(s)))ds\right|
\end{equation*}
\begin{equation}
\leq \int_{0}^{\xi +\epsilon }\left| \frac{\partial c}{\partial x}(\zeta
_{n}(t,Z),U_{t,n}(V_{t,n}(s))d(\frac{Z-G_{t}^{-1}(V_{t,n}(s))}{Z})\right.
\label{method2}
\end{equation}
\begin{equation*}
\left. -\frac{\partial c}{\partial x}(G_{t}(Z),s)d(\frac{Z-G_{t}^{-1}(s)}{Z}%
)\right| e(G_{t}^{-1}(V_{t,n}(s)))ds
\end{equation*}
This latter tends uniformly in $\phi $ to zero since $\frac{\partial c}{%
\partial x}(\cdot ,\cdot )$ is uniformly continuous on $\left[ 0,\xi
+\epsilon \right] $ in the sense described in (HP3). By letting $\epsilon
\rightarrow 0$, we get the convergence of \textcolor{blue}{(\ref{method2})}
to zero, uniformly in $\phi ,$ in outer probability. Remark further that 
\begin{equation*}
\left| \int_{0}^{1}\frac{\partial c}{\partial x}(G_{t}(Z),s)d(\frac{%
Z-G_{t}^{-1}(s)}{Z})e(G_{t}^{-1}(V_{t,n}(s)))ds-K_{c}\right|  \label{method3}
\end{equation*}

\begin{equation}
\leq \int_{0}^{1}\left| \frac{\partial c}{\partial x}(G_{t}(Z),s)d(\frac{%
Z-G_{t}^{-1}(s)}{Z})\right| \left|
e(G_{t}^{-1}(V_{t,n}(s)))-e(G_{t}^{-1}(s))\right| ds.  \label{method4}
\end{equation}
Now $\left| e(G_{t}^{-1}(V_{t,n}(s)))-e(G_{t}^{-1}(s))\right| $ is the
indicator function of the symetrical difference of the sets $(V_{t,n}(s)\leq
G(Z))$ and $(s\leq G_{t}(Z))$. And we have, for large values of n, $%
w.p.1-\epsilon $

\begin{equation}
(V_{t,n}(s)\leq G_{t}(Z))\Delta (s\leq G_{t}(Z))\subseteq  \label{method5}
\end{equation}
\begin{equation*}
(G_{t}(Z)-\epsilon \leq s<G_{t}(Z))+(G_{t}(Z)\leq s\leq G_{t}(Z)+\epsilon ).
\end{equation*}
This and the uniform boundedness in $\phi ,$ (say by M), of the functions $%
\frac{\partial c}{\partial x}(G_{t}(Z),s)d(\frac{Z-G_{t}^{-1}(s)}{Z})$, due
its uniform continuity on $\left[ 0,G(Z)+\epsilon _{0}\right] $ in the sense
of (HP3), imply that the second member in \textcolor{blue}{(\ref{method4})}
is, uniformly in $\phi ,$ less than $2M\epsilon $, $w.p.1-\epsilon $. By
letting $\epsilon \rightarrow 0$, we finally get the convergence of the
integral in \textcolor{blue}{(\ref{method4})}, uniformly in $\phi ,$ to $K_{c}$. This kind of arguments will be used in the sequel whithout further
details. It follows that 
\begin{equation}
D_{1}=K_{c}(G_{n}(Z)-G(Z))/\sqrt{n}+o_{p}^{\ast }(n^{-1/2}),  \label{method6}
\end{equation}
uniformly in $\phi .$ Next 
\begin{equation}
D_{2}=\frac{1}{n}\sum \left\{ G_{t,n}(Y_{j}(t))-G_{t}(Y_{j}(t))\right\} 
\frac{\partial c}{\partial y}(G_{t}(Z),\zeta _{t,n}(j))\text{ }\gamma
(Y_{j}(t));  \label{method8}
\end{equation}
where $\zeta _{t,n}(j)$ lies between $G_{t,n}(Y_{j}(t))$ and $%
G_{t}(Y_{j}(t)).$ Now, denoting $I_{n}=[G_{t,n}(Y_{j}(t))\wedge
G_{t}(Y_{j}(t)),G_{t,n}(Y_{j}(t))\vee G_{t}(Y_{j}(t))],$ we have to show
that 
\begin{equation}
\max_{1\leq j\leq n}\sup_{t\in \lbrack 0,T]}\sup_{(x,y)\in I_{n}^{2}}\left| 
\frac{\partial c}{\partial y}(G_{t}(Z),x)-\frac{\partial c}{\partial y}%
(G_{t}(Z),y))\text{ }\right| \gamma (Y_{j}(t))\rightarrow _{P^{\ast }}0,
\label{method9}
\end{equation}
as $n\rightarrow +\infty$. The boundedness of $\gamma (\cdot )$, the equi-continuity of $\frac{\partial
c}{\partial y}$ and (HP1) establish \textcolor{blue}{(%
\ref{method9})}. By $(HP7)$ and Theorem 2 in \textcolor{red}{\cite{lo1}},
the process $B_{n}^{\ast }(\cdot)$ defined by: 
\begin{equation*}
\left\{ B_{n}^{\ast }(t),0\leq t\leq T\right\} =\left\{ \frac{1}{\sqrt{n}}%
\sum \left\{ G_{t,n}(Y_{j}(t))-G_{t}(Y_{j}(t))\right\} ,0\leq t\leq T\right\}
\end{equation*}
converges to a Gaussian process $\{G_{0}(t),0\leq t\leq T\}$ in $\ell
^{\infty }([0,T])$ under the hypotheses. Thus, 
\begin{equation*}
\sqrt{n}D_{2}=\frac{1}{\sqrt{n}}\sum \left\{
G_{t,n}(Y_{j}(t))-G_{t}(Y_{j}(t))\right\} \frac{\partial c}{\partial y}%
(G_{t}(Z),G_{t}(Y_{j}(t))\text{ }\gamma (Y_{j}(t))+o_{P}^{\ast }(1),
\end{equation*}
uniformly in $\phi \in \Phi .$ We conclude that 
\begin{equation*}
\sqrt{n}(J_{n}(c)-H_{c})=\frac{1}{\sqrt{n}}\sum g(Y_{j})-Eg(Y_{j})+K_{1}%
\sqrt{n}(G_{t,n}(Z)-G(Z)
\end{equation*}
\begin{equation*}
+\frac{1}{\sqrt{n}}\sum \left\{ G_{t,n}(Y_{j}(t))-G_{t}(Y_{j}(t))\right\}
\nu _{c,t}(y)+o_{P}^{\ast }(1)
\end{equation*}

\bigskip

\noindent Now, we have to handle in the same lines to get for $J_{n}(\pi)$
\begin{equation*}
\sqrt{n}(J_{n}(\pi )-H_{\pi })=\frac{1}{\sqrt{n}}\sum g_{\pi
}(Y_{j})-Eg_{\pi }(Y_{j})+K_{\pi }\sqrt{n}(G_{t,n}(Z)-G(Z)
\end{equation*}%
\begin{equation*}
+\frac{1}{\sqrt{n}}\sum \left\{ G_{t,n}(Y_{j}(t))-G_{t}(Y_{j}(t))\right\}
\nu _{\pi ,t}(Y_{j})+o_{P}^{\ast }(1).
\end{equation*}%

\noindent Remark that $H_{c}(\phi )$ and $H_{\pi }(\phi )$ are uniformly bounded. Then we arrive at the representation (R).\\

\noindent Now, when $(HP2c)$ holds for $h(n,Q)=B(n,Q)$, the quotient of \textcolor{blue}{(\ref{p01})} is one. This leads to the representation (RD) only based on that of $J_{n}(c)$.

\newpage

\end{document}